%% file: main.tex
\documentclass[10pt,journal,compsoc]{IEEEtran} 

\usepackage{amsmath,amssymb,amsfonts,amsthm}
\usepackage{algorithm}
\usepackage{algorithmic}
\usepackage{graphicx}
\usepackage{subfigure}
\usepackage{cite}

\usepackage{booktabs}
\usepackage{multicol}
\usepackage{multirow}
\usepackage{threeparttable}

\usepackage{xcolor}
\usepackage{url}
\usepackage[hidelinks]{hyperref}
\usepackage{eso-pic}

\AddToShipoutPictureBG*{%
  \AtPageLowerLeft{%
    \raisebox{1.0cm}{%
      \makebox[\paperwidth][c]{%
        \parbox{0.86\paperwidth}{%
          \footnotesize \centering
          \textcopyright~2024 IEEE. Personal use of this material is permitted. Permission from IEEE must be obtained for all other uses, in any current or future media, including reprinting/republishing this material for advertising or promotional purposes, creating new collective works, for resale or redistribution to servers or lists, or reuse of any copyrighted component of this work in other works. DOI: \href{https://doi.org/10.1109/TC.2024.3483632}{10.1109/TC.2024.3483632}
        }%
      }%
    }%
  }%
}

\begin{document}

\title{Bit-Sparsity Aware Acceleration with Compact CSD Code on Generic Matrix Multiplication}

\author{
  Zixuan Zhu, 
  Xiaolong Zhou, 
  Chundong Wang,
  Li Tian, 
  Zunkai Huang, ~\IEEEmembership{Member,~IEEE}, \\
  and Yongxin Zhu, ~\IEEEmembership{Senior Member,~IEEE}%
  \thanks{This work was supported in part by the National Key Research and Development Program of China under Grant 2023YFB2807001, in part by Shanghai Talent Development Fund under Grant E1322E1, in part by the National Ministry of Industry and Information Technology High-Quality Development Project of China under Grant E412151, in part by the Research on Key Technologies of AI-Based Energy Spectrum Detection Electronics Equipment under Grant E2520F1, and in part by the National SKA Program of China under Grant 2020SKA0120202.}
}

\markboth{}%
{}%

\maketitle

\input{tex/Abstract}

\begin{IEEEkeywords}
  Bit-level sparsity, Canonical signed digit, Hardware acceleration, Floating point, Matrix multiplication.
\end{IEEEkeywords}

\input{tex/Introduction.tex}
\input{tex/Background.tex}
\input{tex/CompactCSD.tex}
\input{tex/Segmenting.tex}
\input{tex/Architecture.tex}
\input{tex/Evaluation.tex}
\input{tex/Sensitivity.tex}
\input{tex/ReatedWork.tex}
\input{tex/Conclusion.tex}

\bibliographystyle{IEEEtran}
\bibliography{refs.bib}

\vfill

\end{document}

%% file: tex/Abstract.tex
\begin{abstract}
The ever-increasing demand for matrix multiplication in artificial intelligence (AI) and generic computing emphasizes the necessity of efficient computing power accommodating both floating-point (FP) and quantized integer (QINT).
While state-of-the-art bit-sparsity-aware acceleration techniques have demonstrated impressive performance and efficiency in neural networks through software-driven methods such as pruning and quantization, these approaches are not always feasible in typical generic computing scenarios.
In this paper, we propose Bit-Cigma, a hardware-centric architecture that leverages bit-sparsity to accelerate generic matrix multiplication.
Bit-Cigma features (1) CCSD encoding, an optimized on-chip sparsification technique based on canonical signed digit (CSD) representation; (2) segmented dot product, a multi-stage exponent matching technique for long FP vectors; and (3) the versatility to efficiently process both FP and QINT data types.
CCSD encoding halves the cost of CSD encoding while achieving optimal bit-sparsity, and segmented dot product improves both accuracy and throughput.
Bit-Cigma cores are implemented using 65 nm technology at 1 GHz, demonstrating substantial gains in performance and efficiency for both FP and QINT configurations.
Compared to state-of-the-art Bitlet, Bit-Cigma achieves 3.2× performance, 6.1× area efficiency, and 15.3× energy efficiency when processing FP32 data while ensuring zero computing error.
\end{abstract}
  

%% file: tex/Introduction.tex
\section{Introduction}\label{sec-introduction}

\IEEEPARstart{M}{atrix} multiplication is a fundamental operation in artificial intelligence (AI) and general-purpose computing, such as novel generative AI \cite{Palm2023, llama2023, Sauer2023}, computer vision techniques \cite{SAM2023}, etc.
To accommodate diverse workloads, various data types are utilized.
While quantized integer (QINT) has proven capability in deep neural network (DNN) inference, floating-point (FP) remains crucial for model training and applications beyond DNN due to its large dynamic range.
This highlights the need for high-performance and efficient FP matrix multiplication accelerators.

Substantial bit-level sparsity in DNN matrices has prompted the development of bit-sparsity-aware acceleration techniques \cite{Balance2024, BitHist2023, Mortar2023}.
These techniques employ bit-serial \cite{Jing2023, Tactical2019, Pragmatic2017} and bit-interleaving \cite{BitletX2024, BEM2022, Tetris2018} strategies to eliminate operations involving zero bits.
Encoding techniques further enhance accelerator performance and efficiency.
For instance, Laconic \cite{Laconic2019} reduces digit count using radix-4 Booth code, while DWP \cite{DWP2022} minimizes non-zero digits through Canonical Signed Digit (CSD) representation, as demonstrated in \cite{HWP2004, Hewlitt2000}.
However, these methods are tailored to QINT DNNs and rely on software re-coding, limiting their applicability in scenarios requiring FP data or prohibiting modifications.

Recently, hardware-centric approaches for bit-sparsity-aware FP acceleration have been explored to eliminate software preliminaries and generalize accelerator use.
Examples include Bitlet \cite{Bitlet2021} and BEM \cite{BEM2022} processing elements (PEs), which accelerate original FP DNNs without software-level modifications.
However, their performance and efficiency are constrained by exponent preprocessing, which involves caching and traversing entire operand vectors.
Moreover, these methods introduce computational errors by discarding essential mantissa bits during exponent matching.
Thus, improving the performance, accuracy, and efficiency of bit-sparsity-aware FP acceleration remains a significant challenge.

This paper explores the hardware-centric solution for generic matrix multiplication by leveraging bit-sparsity with CSD representation.
Most non-DNN applications cannot adopt software modification techniques, and FP data types are inherently represented in binary formats.
This necessitates an efficient on-chip encoding method.
Additionally, existing exponent-matching strategies introduce significant memory and data access overhead, prevent the use of high-throughput techniques, and lead to ineluctable computational errors, highlighting the need for a more efficient exponent-matching approach.

One challenge, regarding CSD application, lies in developing a cost-efficient and fast encoder circuit.
Although CSD with minimal non-zero digits is optimal for bit-sparsity acceleration, existing conversion techniques hinder on-chip deployment performance and efficiency.
Current methods for converting N-bit 2's complement to CSD codes \cite{Tanaka2016, Ruiz2011, Guo2011} require 2N-bit memory and an N-bit carry chain, but their large area and propagation delay limit on-chip re-coding speed and efficiency.
While software conversion in DWP \cite{DWP2022} reduces memory requirements by regrouping CSD digits, this approach is unsuitable for circuit implementation and general use.

Another challenge, regarding FP computing, is achieving efficient dynamic exponent matching without sacrificing accuracy.
Previous approaches have shown significant overheads in exponent matching.
For example, Bitlet \cite{Bitlet2021} must stall to wait for exponent matching and requires large memory to cache operands.
An alternative is Flex, a custom FP vector datatype \cite{NeuroFlix2022, FlexBlock2023}, which stores a group of aligned mantissas with a shared exponent to avoid on-chip dynamic exponent matching.
However, the limited value diversity within a Flex vector may restrict its applicability compared to standard FP formats.

To address these challenges, we propose Bit-Cigma, a \underline{C}ompact CSD (CCSD) \underline{I}nspired \underline{G}eneric \underline{M}atrix multiplication \underline{A}rchitecture for \underline{Bit}-sparsity-aware FP acceleration.
At the data level, the CCSD encoding technique enables on-chip bit-level sparsification.
At the PE level, preprocessors and vector PEs exploiting CCSD sparsity enhance throughput and accuracy by segmented dot products.
Additionally, we implement and evaluate a degenerated version of QINT.
With the CCSD technique, Bit-Cigma employs standard FP or QINT formats for memory access, converting data within the core to boost performance without incurring additional memory access overhead.

In summary, this paper focuses on bit-sparsity-aware acceleration for generic matrix multiplication, with the following key contributions:
\begin{itemize}
  \item We propose the Bit-Cigma architecture, which features active on-chip sparsification, flexible datatype support, and highly accurate generic matrix multiplication.
  \item We design a CSD-based on-chip conversion technique, termed CCSD encoding, that achieves optimal bit-sparsity with compact and fast encoder circuits.
  \item We put forward a segmented approach for FP exponent matching and dot product computation, improving vector PE throughput.
\end{itemize}

Specifically, Bit-Cigma significantly improves performance, efficiency, and accuracy through CCSD encoding and segmented dot products.
For FP32 acceleration, Bit-Cigma outperforms the state-of-the-art Bitlet \cite{Bitlet2021} by 3.2× in speedup ratio, 6.1× in area efficiency, and 15.3× in energy efficiency.
Moreover, Bit-Cigma eliminates FP computing errors.
For QINT16 acceleration, the degenerated Bit-Cigma achieves 2.4× performance and 3.4× area efficiency compared to the state-of-the-art Tactical \cite{Tactical2019}.

The remainder of this paper is organized as follows:
Sec. \ref{sec-background} provides background and motivation.
Sec. \ref{sec-ccsd} details our CCSD code and encoder schemes, Sec. \ref{sec-segmenting} describes the segmented approach, and Sec. \ref{sec-architecture} presents the proposed Bit-Cigma design.
Sec. \ref{sec-evaluation} evaluates the performance and efficiency of Bit-Cigma implementations, with sensitivity studies on key design parameters in Sec. \ref{sec-sensitivity}.
Related work on bit-sparsity-aware acceleration and CSD encoding techniques are discussed in Sec. \ref{sec-relatedwork}, and conclusions are drawn in Sec. \ref{sec-conclusion}.

%% file: tex/Background.tex
\section{Background and Motivation}\label{sec-background}

\begin{figure*}[t]
  \centering
  \includegraphics[scale=0.8]{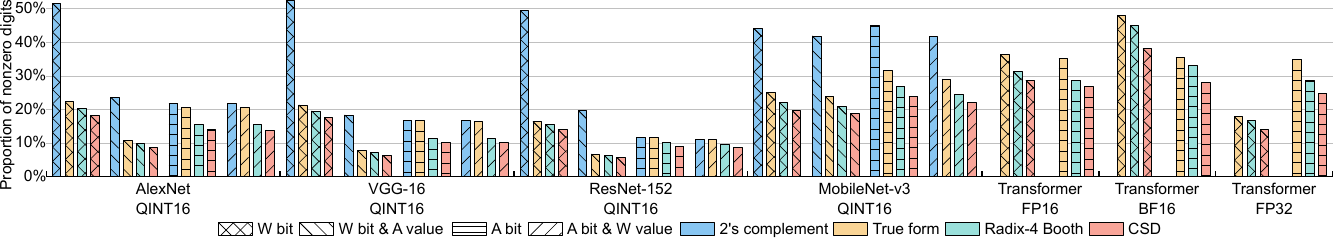}
  \caption{Proportion of non-zero digits in bit-matrices from neural networks converted to INT16/Q12, FP16, BF16, and FP32 with various bit-sparsity-aware strategies and encoding methods. Lower is better. Two key comparisons are highlighted: (1) density reduction using different strategies (by pattern), and (2) density reduction using different encoding methods (by color).}
  \label{fig-motiv-CodeDensity}
\end{figure*}

\subsection{FP Bit-Sparsity Acceleration}\label{subsec-background-sparsity}

Bit-sparsity acceleration eliminates operations for zero partial products.
Consider an integer dot product where vectors $\mathbf{W}$ and $\mathbf{A}$ each contain $N$ items, with $K$ bits per item.
At the bit level, the dot product can be expressed as:
\begin{equation}
  \begin{aligned}
    \mathbf{W} \cdot \mathbf{A} &= \sum_{n}^{N}{W_n \times A_n} = \sum_{n}^{N}\sum_{k}^{K}{P_{n,k}} \Bigg|_{P_{n,k} \neq 0} \\
    P_{n,k} &= w_{n,k} \times A_n \times 2^k
  \end{aligned}
  \label{eq-qintmatbit}
\end{equation}
where $P_{n,k}$ is a partial product, $W_n$ and $A_n$ are the $n$-th items of operand vectors, and $w_{n,k}$ is the bit of $W_n$ with significance $2^k$.
$\mathbf{W}$ and $\mathbf{A}$ distinguish two operand vectors, not necessarily weights or activations in DNN hereafter.

An FP value is typically represented as:
$$
\text{val} = {(-1)}^S \times 2^E \times M
$$
where $S$ is the sign (0 for positive, 1 for negative), $E$ is the exponent (without bias), and $M$ is the fixed-point mantissa.
Since exponents typically differ, direct summation of mantissas is impractical.
A common approach is to align all operands to the maximum exponent.

The dot product of $N$-item FP vectors, assuming a mantissa width of $K$, can be expressed as:
\begin{equation}
  \begin{aligned}
    \mathbf{W} \cdot \mathbf{A} &= 2^{E_\text{max}} \times \sum_{n}^{N}\sum_{k}^{K}{ P_{n,k}} \Bigg|_{P_{n,k} \neq 0 } \\
    P_{n,k} &= {\left(-1\right)}^{W_n^S \oplus A_n^S} \times w_{n,k} \times A_n^M \times 2^{\left(k+E_n^\text{sum}-E_\text{max}\right)} \\
    E_n^\text{sum} & = W_n^E+A_n^E \\
    E_\text{max} &= \max{\left\{E_n^\text{sum}\right\}}
  \end{aligned}
  \label{eq-floatmac}
\end{equation}
where $W_n^S$, $A_n^S$, $W_n^E$, $A_n^E$, $W_n^M$, and $A_n^M$ represent the sign, exponent, and mantissa for the $n$-th item of the operand vectors.
The operator $\oplus$ denotes logical XOR.
$E_n^\text{sum}$ is the exponent sum of the $n$-th item pair, and $E_\text{max}$ is the maximum exponent sum among all item pairs.

Traditional multipliers generate and sum $N \times K$ partial products, whereas bit-sparsity accelerators generate partial products only when $w_{n,k} \neq 0$.
Zero items in $\mathbf{A}$ can also eliminate unnecessary $P_{n,k}$.
This leads to four bit-sparsity-aware acceleration properties: weight bit-sparsity (W bit), weight bit-sparsity with activation value-sparsity (W bit \& A value), activation bit-sparsity (A bit), and activation bit-sparsity with weight value-sparsity (A bit \& W value).

As shown in Fig. \ref{fig-motiv-CodeDensity}, the impact of sparsity-aware properties varies across models.
When changing the property, some DNNs, like VGG-16 \cite{Vgg2015} and ResNet-152 \cite{Resnet2016}, show significant reductions in non-zero digits, while others, such as MobileNet-v3 \cite{MobileNet2019} and Transformer \cite{Transformer2017}, do not.

\subsection{Canonical Signed Digit Representation}\label{sec-background-csd}

The Canonical Signed Digit (CSD) representation, also known as the Non-Adjacent Form (NAF), is a unique signed digit (SD) representation with the digit set $\{\bar{1}~0~1\}$.
Here, a digit with a bar means a negative digit, $\bar{1}$ for -1 and $\bar{2}$ for -2 hereafter.
CSD has two key properties essential for on-chip application: the minimal Hamming weight (\textbf{sparsest property}) and the non-adjacency of any two non-zero digits (\textbf{non-adjacent property}).
The sparsest property implies minimal non-zero digits, while the non-adjacent property distinguishes CSD from other minimal signed digit (MSD) representations that satisfy the sparsest property.

Consider $(-23)_{10}$ as an example.
Tab. \ref{tab-motiv-SDCodes} lists several SD representations commonly used in accelerator design, along with their corresponding Hamming weights.
CSD consistently exhibits the lowest count of non-zero digits.

\begin{table}[t]
  \centering
  \caption{Signed digit representations and Hamming weights.}
  \label{tab-motiv-SDCodes}
  \scriptsize
  \input{tab/tab-motiv-SDCodes.tex}
\end{table}

Theoretically, CSD contains $1/3$ fewer non-zero digits on average than 2's complement and true form (see Tab. \ref{tab-motiv-SDCodes}).
For an $N$-bit integer, 2's complement, true form, and radix-2 Booth code typically contain $N/2$ non-zero digits on average, with a maximum of $N$.
In contrast, MSD representations, including CSD, contain $N/3$ non-zero digits on average and a maximum of $N/2$.
The reduced number of non-zero digits in CSD inspires more efficient bit-sparsity-aware acceleration, as converting operands to CSD reduces the number of essential partial products without accuracy loss, leading to substantial performance gains.

The encoding comparison (by color) in Fig. \ref{fig-motiv-CodeDensity} shows that CSD consistently achieves minimal non-zero digits and the best bit-level sparsity.
With the W bit \& A value property, CSD reduces non-zero digits by approximately $2/3$ compared to 2's complement, by at least 15\% compared to true form, and by about 10\% compared to radix-4 Booth code.

\subsection{Inefficiencies in On-Chip CSD Application}\label{sec-motiv-redundency}

The non-adjacency property of CSD ensures no more than $N/2$ non-zero digits in an $N$-digit CSD code.
Since bit-sparsity accelerator resource and performance are sensitive to data width, redundant zero digits in CSD code limit the potential benefits of on-chip CSD application in encoding, storage, and zero-skip checking:

(1) \textbf{Encoding.}
Prior works \cite{Ruiz2011, Guo2011, Tanaka2016} use an $N$-bit carry chain for $N$-digit CSD code, leading to critical propagation delays for generating redundant zero digits.

(2) \textbf{Storage.}
CSD demands $2N$ bits to store $N$ digits for an $N$-bit integer, doubling the memory and data access bandwidth requirements.

(3) \textbf{Zero-skip checking.}
The check window size, or the number of bits available to the zero-skip logic, is crucial for accelerator performance.
But it is underutilized when checking $N$ bits to find at most $N/2$ non-zeros.

We propose the CCSD code with a holistic encoding approach to eliminate these inefficiencies and support on-chip CSD applications (detailed in Sec. \ref{sec-ccsd}).
CCSD encodes an $N$-bit integer using $N/2$ digits.
This allows the CCSD encoder to halve the carry chain length and reduces the memory requirement to $3N/2$ bits.
Also, with halved digits, the effective check window size is doubled, further enhancing the efficiency of the zero-skip logic.

\subsection{Challenges in FP Bit-Sparsity Acceleration}\label{subsec-background-challenges}

FP dot products require exponent matching, which introduces challenges in data access, accuracy, and performance:

(1) \textbf{Data access.}
Searching for the maximum exponent requires accessing all data before processing, leading to duplicated memory access or large buffers in PE.

(2) \textbf{Accuracy.}
Exponent matching may discard essential bits due to right-shifting, causing accuracy degradation.

(3) \textbf{Performance.}
The need for the maximum exponent forces the processor to stall until the entire operand vector is processed, reducing throughput.

To address these challenges, we propose a segmented approach (detailed in Sec. \ref{sec-segmenting}) and the Bit-Cigma architecture for FP bit-sparsity acceleration (discussed in Sec. \ref{sec-architecture}).
The segmented dot product method processes only part of the operand vectors, avoiding accessing large amounts of data and pausing to wait for data.
Exponent matching is postponed until summation, allowing for improved accuracy by adjusting arithmetic data width and enhanced throughput by simultaneous zero-bit skipping and exponent matching.

%% file: tab/tab-motiv-SDCodes.tex
\begin{threeparttable}
\begin{tabular}{rcccc}
\toprule
\multirow{2}{*}{Representation} & \multicolumn{2}{c}{Encoding $(-23)_{10}$} & \multicolumn{2}{c}{Encoding N-bit int} \\
\cmidrule(lr){2-3} \cmidrule(lr){4-5}
                       & SD code                              & HW\tnote{1} & Avg. HW & Max. HW \\
\midrule
        2's complement & $\bar{1}110\_1001$                   & 5           & N/2     & N      \\
             True form & $000\bar{1}\_0\bar{1}\bar{1}\bar{1}$ & 4           & N/2     & N      \\
         Radix-2 Booth & $00\bar{1}1\_\bar{1}01\bar{1}$       & 5           & N/2     & N      \\
Radix-4 Booth\tnote{*} & $000\bar{1}\_\bar{1}001$             & 3           & 3N/8    & N/2    \\
                   CSD & $00\bar{1}0\_1001$                   & 3           & N/3     & N/2    \\
             other MSD & $000\bar{1}\_\bar{1}001$             & 3           & N/3     & N/2    \\
\bottomrule
\end{tabular}
\begin{tablenotes}
  \item[*] Converted from digit set $\{\bar{2}~\bar{1}~0~1~2\}$ to $\{\bar{1}~0~1\}$ to compare.
  \item[1] Hamming weight, the quantity of non-zero digits in SD code.
\end{tablenotes}
\end{threeparttable}

%% file: tex/CompactCSD.tex
\section{Compact CSD Encoding}\label{sec-ccsd}

Regarding efficient on-chip CSD application, we design the Compact CSD (CCSD) code and encoder.
Sec. \ref{sec-ccsd-code} presents the CCSD code and its binary record.
Sec. \ref{sec-ccsd-encoder} elaborates on the CCSD encoding methods and encoder circuits.
Sec. \ref{sec-ccsd-proof} provides proof of the CCSD encoding technique.

\begin{table}[t]
  \centering
  \caption{Encoding map from CSD to CCSD.}
  \label{tab-ccsd-encodemap}
  \scriptsize
  \input{tab/tab-ccsd-EncodeMap.tex}
\end{table}

\subsection{Compact CSD Code}\label{sec-ccsd-code}

The non-adjacency property of CSD representation implies specific combinations for even-odd digit pairs, i.e., $\bar{1}0$, $10$, $00$, $01$, and $0\bar{1}$.
CCSD merges each even-odd digit pair and represents integers using digit set $\{\bar{\text{2}}~\bar{\text{1}}~\text{0}~\text{1}~\text{2}\}$.
Thus, an $N$-digit CSD code (where $N$ is even) is converted into an $N/2$-digit CCSD code, with each CCSD digit requiring 3-bit storage.
This effectively halves the digit number and reduces the memory space to $3N/2$ bits per value.

To record CCSD digits in binary, we employ a group of code marks, including a 1-bit sign mark $E^s_k$, a 1-bit double mark $E^l_k$, and a 1-bit value mark $E^v_k$.
Specifically, $E^v_k = 1$ indicates a non-zero digit, while $E^v_k = 0$ indicates a zero digit.
For non-zero digits, $E^s_k = 1$ signifies a negative digit, whereas $E^s_k = 0$ signifies a positive digit.
Additionally, $E^l_k = 1$ represents $\pm 2$ digits, while $E^l_k = 0$ represents $\pm 1$ digits.
When $E^v_k = 0$, all combinations of $E^s_k$ and $E^l_k$ represent the digit zero.

As shown in the encoding map in Tab. \ref{tab-ccsd-encodemap}, the value of a CCSD digit is given by:
$$
E_k = {\left(-1\right)}^{E_k^s} \times 2^{E_k^l} \times E_k^v
$$
and the value of a $K$-digit CCSD code is expressed as:
$$
\text{val} = \sum_{k=0}^{K-1}{\left(E_k \times 4^k\right)} = \sum_{k=0}^{K-1}{\left({\left(-1\right)}^{E_k^s} \times 2^{2k+E_k^l}\right)} \Bigg|_{E_k^v \neq 0}
$$

This binary representation serves two purposes.
First, it can be directly utilized for arithmetic operations without further decoding, where $E^s_k = 1$ triggers an inversion, $E^l_k = 1$ denotes doubling the operand, and $E^v_k = 0$ indicates a zero-skip.
Second, this representation facilitates an efficient encoding process elaborated upon in Sec. \ref{sec-ccsd-encoder}.

\begin{table}[t]
  \centering
  \caption{Truth table for the proposed CCSD encoding method.}
  \label{tab-ccsd-truthtable}
  \scriptsize
  \input{tab/tab-ccsd-TruthTable.tex}
\end{table}

\begin{figure}[t]
  \centering
  \includegraphics{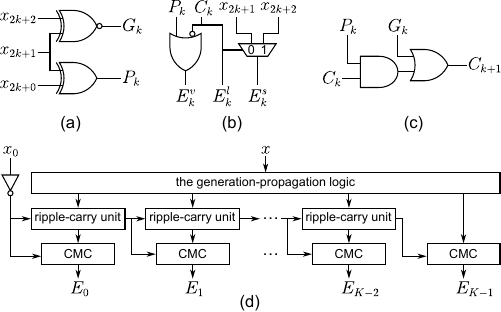}
  \caption{Components and schemes of the proposed CCSD encoder. (a) The generation-propagation logic. (b) The code mark converter (CMC). (c) The basic ripple-carry unit. (d) A $K$-digit CCSD ripple-carry encoder.}
  \label{fig-ccsd}
\end{figure}

\subsection{CCSD Encoders}\label{sec-ccsd-encoder}

To convert 2's complement into CCSD binary records, we propose an encoding technique based on
the generation-propagation logic:
\begin{subequations}
  \begin{align}
    P_k & = x_{2k+1} \oplus x_{2k+0} \label{eq-pglogic-p} \\
    G_k & = x_{2k+2} \odot x_{2k+1} \label{eq-pglogic-g}
  \end{align}
  \label{eq-pglogic}
\end{subequations}
the ripple carry chain logic:
\begin{subequations}
  \begin{align}
    C_{k+1} & = G_k + P_k C_k \label{eq-carrychain-k} \\
    C_0     & = \overline{x_0} \label{eq-carrychain-0}
  \end{align}
  \label{eq-carrychain}
\end{subequations}
and the code mark generation logic:
\begin{subequations}
  \begin{align}
    E^s_k & = \begin{cases}
                x_{2k+1}, & C_k = 0 \\
                x_{2k+2}, & C_k = 1
              \end{cases} \label{eq-ccsdcodegen-s} \\
    E^l_k & = C_k \label{eq-ccsdcodegen-d}         \\
    E^v_k & = P_k + \overline{C_k} \label{eq-ccsdcodegen-v}
  \end{align}
  \label{eq-ccsdcodegen}
\end{subequations}

In these equations, $\oplus$ denotes logical XOR, and $\odot$ denotes logical XNOR.
The variable $x_i$ represents the $i$-th bit of the sign-extended 2's complement.
The encoding truth table from 2's complement to CCSD binary records is detailed in Tab. \ref{tab-ccsd-truthtable}.
Inputs include $x_{2k+2}$, $x_{2k+1}$, and $x_{2k+0}$ from 2's complement, while outputs are $E^s_k$, $E^l_k$, and $E^v_k$ for the CCSD code, and $C_k$, $C_{k+1}$ as carry bits.

The CCSD ripple-carry encoder is derived from Eq. \ref{eq-pglogic}, Eq. \ref{eq-carrychain}, and Eq. \ref{eq-ccsdcodegen}.
Since the ripple-carry encoder shares the carry chain expression (Eq. \ref{eq-carrychain-k}) with the ripple-carry adder, all carry chain optimization techniques that benefit from the generation-propagation logic, such as carry-look-ahead, prefix, carry-select, etc., can be directly applied to CCSD encoders.
This design also allows compilers to detect the carry chain, enabling further optimization.

The components and schemes of the CCSD encoder are illustrated in Fig. \ref{fig-ccsd}.
A CCSD encoder consists of three parts: the generation-propagation logic (Fig. \ref{fig-ccsd}(a)), the code mark converter (Fig. \ref{fig-ccsd}(b)), and a carry chain with specific optimization.
Fig. \ref{fig-ccsd}(c) shows the basic ripple-carry unit, and Fig. \ref{fig-ccsd}(d) depicts a $K$-digit ripple-carry encoder.

\subsection{Proof}\label{sec-ccsd-proof}

The proof of the CCSD encoder, derived from Eq. \ref{eq-pglogic}, Eq. \ref{eq-carrychain}, and Eq. \ref{eq-ccsdcodegen}, is as follows.

In prior work \cite{Guo2011}, 2's complement is converted into CSD sign marks and magnitude marks using:
\begin{subequations}
  \begin{align}
    D^m_i & = \begin{cases}
                \left(x_i \oplus x_{i-1}\right) \overline{D^m_{i-1}}, & i > 0 \\
                x_0,                                                  & i = 0
              \end{cases} \label{eq-bin2csd-m} \\
    D^s_i & = x_{i+1} \label{eq-bin2csd-s}
  \end{align}
  \label{eq-bin2csd}
\end{subequations}
where $D^s_i$ and $D^m_i$ are the $i$-th CSD sign mark and magnitude mark, respectively.

Letting the CCSD carry bit $C_k = \overline{D^m_{2k}}$ and applying the generation-propagation logic (Eq. \ref{eq-pglogic}), we deduce the CCSD carry chain as
$$
\begin{aligned}
  C_{k+1} & = \overline{D^m_{2k+2}}                                                                            \\
          & = \overline{\left(x_{2k+2} \oplus x_{2k+1}\right) \overline{D^m_{2k+1}}}                           \\
          & = \overline{\left(x_{2k+2} \oplus x_{2k+1}\right)} + D^m_{2k+1}                                    \\
          & = \left(x_{2k+2} \odot x_{2k+1}\right) + \left(x_{2k+1} \oplus x_{2k+0}\right) \overline{D^m_{2k}} \\
          & = G_k + P_k C_k                                                                                    \\
  C_0     & = \overline{D^m_0} = \overline{x_0}
\end{aligned}
$$

From Tab. \ref{tab-ccsd-encodemap}, CSD code marks can be converted into the proposed CCSD code marks by
\begin{subequations}
  \begin{align}
    E^s_k & = \begin{cases}
                D^s_{2k+0}, & E^l_k = 0 \\
                D^s_{2k+1}, & E^l_k = 1
              \end{cases} \label{eq-csd2ccsd-s}             \\
    E^v_k & = D^m_{2k+1} + D^m_{2k+0} \label{eq-csd2ccsd-v} \\
    E^l_k & = \overline{D^m_{2k+0}} \label{eq-csd2ccsd-d0}  \\
    E^l_k & = D^m_{2k+1} \label{eq-csd2ccsd-d1}
  \end{align}
  \label{eq-csd2ccsd}
\end{subequations}
Note that the $k$-th CCSD double mark $E^l_k$ has two possible forms: Eq. \ref{eq-csd2ccsd-d0} or Eq. \ref{eq-csd2ccsd-d1}.

To replace the CSD code marks with CCSD carry bits, we eliminate $D^m_{2k+1}$ using Eq. \ref{eq-bin2csd-m} to keep only even-indexed terms.
Applying Eq. \ref{eq-bin2csd-s}, the CCSD code mark generation logic is deduced as follows:
$$
\begin{aligned}
  E^l_k & = \overline{D^m_{2k+0}} = C_k \\
  E^s_k & = \begin{cases}
              D^s_{2k+0}, & E^l_k = 0 \\
              D^s_{2k+1}, & E^l_k = 1
            \end{cases} \quad
          = \begin{cases}
              x_{2k+1}, & C_k = 0 \\
              x_{2k+2}, & C_k = 1
            \end{cases} \\
  E^v_k & = D^m_{2k+1} + D^m_{2k} \\
        & = \left(x_{2k+1} \oplus x_{2k+0}\right) \overline{D^m_{2k}} + D^m_{2k} \\
        & = P_k C_k + \overline{C_k} \\
        & = P_k + \overline{C_k}
\end{aligned}
$$

%% file: tab/tab-ccsd-EncodeMap.tex
\begin{threeparttable}
  \begin{tabular}{cccccccccc}
    \toprule
    \multicolumn{3}{c}{Signed digit} & \multicolumn{5}{c}{Binary record\tnote{*}} \\
    \cmidrule(lr){1-3} \cmidrule(lr){4-8}
    \multicolumn{2}{c}{CSD} & CCSD & \multicolumn{2}{c}{CSD\tnote{a}} & \multicolumn{3}{c}{CCSD\tnote{b}} \\
    \cmidrule(lr){1-2} \cmidrule(lr){3-3} \cmidrule(lr){4-5} \cmidrule(lr){6-8}
    [2k+1]    & [2k+0]    & [k]       & ~$D^{s,m}_{2k+1}$~ & ~$D^{s,m}_{2k+0}$~ & $E^s_k$ & $E^l_k$ & $E^v_k$ \\
    \midrule
    1         & 0         & 2         & 0 ~~~~~ 1          & x ~~~~~ 0          & 0       & 1       & 1       \\
    0         & 1         & 1         & x ~~~~~ 0          & 0 ~~~~~ 1          & 0       & 0       & 1       \\
    0         & 0         & 0         & x ~~~~~ 0          & x ~~~~~ 0          & x       & x       & 0       \\
    0         & $\bar{1}$ & $\bar{1}$ & x ~~~~~ 0          & 1 ~~~~~ 1          & 1       & 0       & 1       \\
    $\bar{1}$ & 0         & $\bar{2}$ & 1 ~~~~~ 1          & x ~~~~~ 0          & 1       & 1       & 1       \\
    \bottomrule
  \end{tabular}
  \begin{tablenotes}
    \item[*] x for don't care, either 1 or 0 is acceptable.
    \item[a] $D^s_i$ and $D^m_i$ are the $i$-th CSD sign and magnitude marks, respectively.
    \item[b] $E^s_k$, $E^l_k$, and $E^v_k$ are the $k$-th CCSD sign, double, and value marks, respectively.
  \end{tablenotes}
\end{threeparttable}

%% file: tab/tab-ccsd-TruthTable.tex
\begin{tabular}{cccccccc}
  \toprule
  \multicolumn{3}{c}{$x_{2k+n}$} & \multicolumn{2}{c}{Carry} & \multicolumn{3}{c}{CCSD} \\
  \cmidrule(lr){1-3} \cmidrule(lr){4-5} \cmidrule(lr){6-8}
  +2 & +1 & +0 & $C_{k}$ & $C_{k+1}$ & $E^s_k$ & $E^l_k$ & $E^v_k$ \\
  \midrule
  0  & 0  & 0  & 0        & 1          & 0       & 0       & 1       \\
  0  & 0  & 1  & 0        & 1          & 0       & 0       & 1       \\
  0  & 1  & 0  & 0        & 1          & 1       & 0       & 1       \\
  0  & 1  & 1  & 0        & 0          & 1       & 0       & 1       \\
  1  & 0  & 0  & 0        & 0          & 0       & 0       & 1       \\
  1  & 0  & 1  & 0        & 1          & 0       & 0       & 1       \\
  1  & 1  & 0  & 0        & 1          & 1       & 0       & 1       \\
  1  & 1  & 1  & 0        & 1          & 1       & 0       & 1       \\
  0  & 0  & 0  & 1        & 1          & 0       & 1       & 0       \\
  0  & 0  & 1  & 1        & 1          & 1       & 1       & 1       \\
  0  & 1  & 0  & 1        & 0          & 0       & 1       & 1       \\
  0  & 1  & 1  & 1        & 0          & 1       & 1       & 0       \\
  1  & 0  & 0  & 1        & 0          & 0       & 1       & 0       \\
  1  & 0  & 1  & 1        & 0          & 1       & 1       & 1       \\
  1  & 1  & 0  & 1        & 1          & 0       & 1       & 1       \\
  1  & 1  & 1  & 1        & 1          & 1       & 1       & 0       \\
  \bottomrule
\end{tabular}

%% file: tex/Segmenting.tex
\section{Segmenting High Dimensions}\label{sec-segmenting}

Regarding bit-sparsity FP acceleration for large vectors and matrices, segmented approaches for vector dot product and matrix multiplications are proposed.
Sec. \ref{sec-segmenting-dotproduct} describes the segmented dot product used for exponent matching of long vectors.
Sec. \ref{sec-segmenting-matrix} shows how to block large matrices.

\begin{algorithm}[t]
  \caption{Segmented dot-product with 2-stage FP exponent matching.}
  \label{alg-segm-dotprod}
  \begin{algorithmic}[1]
    \REQUIRE Segments of vector $\mathbf{W}$ and $\mathbf{A}$
    \ENSURE The dot product of vector $\mathbf{W}$ and $\mathbf{A}$
    \STATE Initialize $E_\text{max}^\text{now}$, $E_\text{max}^\text{prev}$, and the accumulator.
    \REPEAT
      \STATE Load a pair of segments and extract FP data.
      \STATE $E_\text{max}^\text{seg} \leftarrow \text{max}(E_n^W + E_n^A)$ for the current segment.
      \STATE $E_\text{max}^\text{prev} \leftarrow E_\text{max}^\text{now}$.
      \STATE $E_\text{max}^\text{now} \leftarrow \text{max}(E_\text{max}^\text{prev}, E_\text{max}^\text{seg})$.
      \STATE Align the accumulation value to $E_\text{max}^\text{now}$.
      \STATE Align all mantissas to $E_\text{max}^\text{now}$.
      \STATE Generate and accumulate partial products.
    \UNTIL{All segments are processed}
    \STATE Convert the result to FP format.
  \end{algorithmic}
\end{algorithm}

\begin{figure*}[t]
  \centering
  \includegraphics{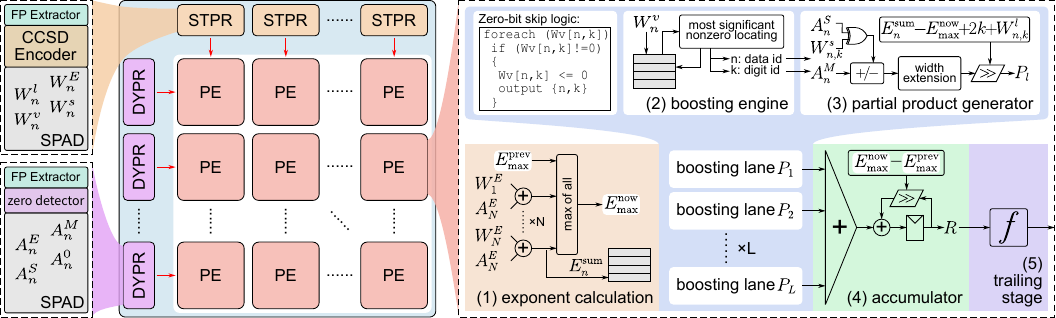}
  \caption{The proposed Bit-Cigma architecture. Peripheries are omitted. Four stages of the PE are distinguished by background colors: orange for the exponent stage, blue for the boosting stage, green for the arithmetic stage, and purple for the trailing stage. The key components of the PE are (1) Exponent calculation block with exponent buffer in the exponent stage, (2) Sparsity boost engine in the boosting stage, (3) Partial product generator in the boosting stage, (4) Multi-data accumulator in the arithmetic stage, and (5) Trailing logic in the trailing stage. Indexes begin from 1.}
  \label{fig-arch}
\end{figure*}

\subsection{Segmented Dot-Product}\label{sec-segmenting-dotproduct}

The traditional method requires the global maximum exponent $E_\text{max}$ to calculate the dot product, as shown in Eq. \ref{eq-floatmac}.
However, in practice, it is expensive and inefficient to obtain $E_\text{max}$ since not all vector items are available.

The segmented dot product is designed to efficiently handle long vectors, especially when not all vector elements are available.
By dividing operand vectors into $N$-item segments, the dot product is calculated as
$$
\mathbf{W} \cdot \mathbf{A} = \left\{\mathbf{W_0}, \mathbf{W_1}, \dots\right\} \cdot \left\{\mathbf{A_0}, \mathbf{A_1}, \dots\right\} = \sum{\left(\mathbf{W_i} \cdot \mathbf{A_i}\right)}
$$
where $\mathbf{W_i}$ and $\mathbf{A_i}$ are segments of $\mathbf{W}$ and $\mathbf{A}$, respectively.
The segmented dot product employs a 2-stage exponent matching process, as outlined in Algorithm \ref{alg-segm-dotprod}.

In the first stage, all mantissas are aligned to the maximum exponent collected from the loaded segments, termed as $E_\text{max}^\text{now}$.
In the second stage, when $E_\text{max}^\text{now}$ is increased in a new segment, the accumulation result is realigned to the updated $E_\text{max}^\text{now}$.
After processing all segments, the final $E_\text{max}^\text{now}$ is the global maximum exponent $E_\text{max}$, and the accumulation result is with $E_\text{max}$.
The accumulation step can incorporate bit-sparsity-aware techniques.

The segmented dot product is also compatible with QINT.
Without exponent-related operations, this procedure degenerates into a multi-input multiplication-accumulation, which does not affect the computing result for QINT. 

\subsection{Blocking the Matrices}\label{sec-segmenting-matrix}

Matrix multiplication is handled by a PE array, where each PE calculates the dot product of a column vector and a row vector.
Large matrices, which exceed the PE array size, are managed as block matrices to fit within the PE array.

For example, the multiplication of matrix $\mathbf{W}$ (with row vectors) and matrix $\mathbf{A}$ (with column vectors) can be expressed as block matrices:
$$
\begin{aligned}
  \left( \begin{array}{c}
           \mathbf{W_0} \\
           \mathbf{W_1} \\
           \mathbf{W_2}
         \end{array} \right) \times
  \left( \begin{array}{cc}
           \mathbf{A_0} & \mathbf{A_1}
         \end{array} \right) =
  \left( \begin{array}{ccc}
    \mathbf{W_0}\!\times\!\mathbf{A_0} & \mathbf{W_0}\!\times\!\mathbf{A_1} \\
    \mathbf{W_1}\!\times\!\mathbf{A_0} & \mathbf{W_1}\!\times\!\mathbf{A_1} \\
    \mathbf{W_2}\!\times\!\mathbf{A_0} & \mathbf{W_2}\!\times\!\mathbf{A_1}
  \end{array} \right)
\end{aligned}
$$
The overall result matrix is obtained by combining the multiplications of the submatrices.

%% file: tex/Architecture.tex
\section{Bit-Cigma Architecture}\label{sec-architecture}

Bit-Cigma leverages the CCSD technique and segmented dot product approach for fast and efficient matrix multiplication.
It is designed for FP but supports QINT as well.
This section provides an overview of the Bit-Cigma architecture (Sec. \ref{sec-architecture-overview}), details the bit-serial vector PE (Sec. \ref{sec-architecture-pe}), and describes the segmented data flow (Sec. \ref{sec-architecture-dataflow}).

\subsection{Overview}\label{sec-architecture-overview}

As illustrated in Fig. \ref{fig-arch}, Bit-Cigma comprises column-shared stationary preprocessors (STPR), row-shared dynamic preprocessors (DYPR), and a PE array.
Data from memory or other PEs, organized as segments, are loaded into the preprocessors, converted, and then transferred to PEs.
The PE array performs bit-sparsity-aware operations, and the results are repacked into the exchange format for output.

Bit-Cigma employs two types of preprocessors for bit-sparsity acceleration.
After extracting data from the exchange formats, the STPR converts values into a CCSD matrix, while the DYPR generates zero marks for the value matrix.
Both preprocessors use scratchpad buffers (SPAD) to cache converted data.
Specifically, DYPR and SPAD utilize ping-pong buffers to enhance throughput.

\subsection{Vector PE}\label{sec-architecture-pe}

The vector PE computes dot products using four stages, as shown in Fig. \ref{fig-arch}.

\textbf{Stage 1: Exponent stage.}
In the exponent stage (the orange part), the exponent block calculates and caches exponent sums in the exponent buffer and determines the maximum exponent for the current segment ($E_\text{max}^\text{now}$).

\textbf{Stage 2: Boosting stage.}
The boosting stage (the blue part) selects non-zero digits from the CCSD matrix and generates corresponding partial products for bit-sparsity acceleration.
Each boosting lane, consisting of a boosting engine and a partial product generator, processes different parts of the bit matrix.
The boosting engine skips zero bits, while the partial product generator produces partial products for non-zero digits.

\textbf{Stage 3: Arithmetic stage.}
The arithmetic stage (the green part) accumulates partial products and performs the second-stage exponent matching.
The multi-data accumulator sums partial products, shifts the accumulation value according to exponents, and integrates the results.
To minimize data transfer in STPRs, the accumulator can store multiple partial results for different dot products.
For example, in $\{\mathbf{W_0},\mathbf{W_1}\} \cdot \{\mathbf{A_0},\mathbf{A_1},\mathbf{A_2}\}$, the accumulator may store three partial results for different $\mathbf{A_i}$, enabling the PE to complete all operations for $\mathbf{W_0}$ before proceeding to $\mathbf{W_1}$.

\textbf{Stage 4: Trailing stage.}
Once all dot-product calculations are completed, the trailing stage (the purple part) processes the output, performing tasks such as activation functions in DNNs, scalar arithmetic in generic matrix multiplications, repackaging results for standard FP formats, etc.

\begin{figure}[t]
  \centering
  \includegraphics{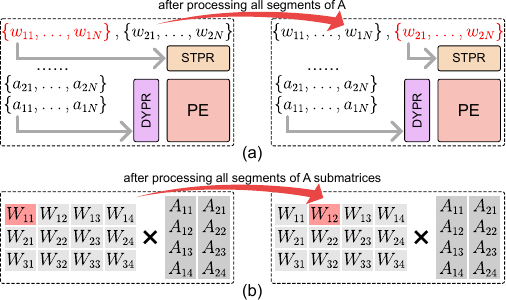}
  \caption{(a) Segmented dot product data flow, where $w_{ij}$ and $a_{ij}$ are elements of vectors. (b) Block matrix data flow, where $W_{ij}$ and $A_{ij}$ are segments of submatrices. Working items of $\mathbf{W}$ are marked in red.}
  \label{fig-dataflow}
\end{figure}

\subsection{Segmented Data Flow}\label{sec-architecture-dataflow}

Each PE handles a dot product of a segment pair from the corresponding STPR and DYPR.
The PE array generates dot products in parallel, contributing to the result matrix for matrix multiplication.
A segmented dot product data flow is designed for matrices within the PE array's capacity, while a block matrix data flow is carried out for larger matrices.
We illustrate matrix multiplication $\mathbf{W} \times \mathbf{A}$ as an example.

\textbf{Segmented dot product data flow.}
As shown in Fig. \ref{fig-dataflow}(a), the segmented dot product data flow proceeds as follows:
(1) Each STPR loads the first segment of a $\mathbf{W}$ row vector.
(2) Each DYPR loads an $\mathbf{A}$ column vector segment-by-segment with ping-pong buffer.
(3) Each PE calculates the dot product of the current segment of $\mathbf{A}$ and $\mathbf{W}$, with partial results stored in the accumulator.
Steps 2 and 3 are performed simultaneously.
(4) After processing all segments of $\mathbf{A}$, the next segment of $\mathbf{W}$ is loaded by the STPR.
Steps 2, 3, and 4 are repeated until all segments of $\mathbf{W}$ are processed.
(5) The trailing stage processes the arithmetic results to generate the final output.

\textbf{Block matrix data flow.}
Fig. \ref{fig-dataflow}(b) illustrates the block matrix data flow.
The process is similar to the segmented dot product data flow, but in step 4, each segment of $\mathbf{W}$ submatrices is processed with multiple $\mathbf{A}$ submatrices before proceeding to the next segment.
For instance, $\mathbf{W_{11}}$ is multiplied with $\mathbf{A_{11}}$ and $\mathbf{A_{12}}$, and then $\mathbf{W_{12}}$ is multiplied with $\mathbf{A_{21}}$ and $\mathbf{A_{22}}$.
Each PE caches an individual partial result for each $\mathbf{A}$ submatrix, producing a row of result submatrices upon completion of each $\mathbf{W}$ submatrix.

%% file: tex/Evaluation.tex
\section{Evaluation}\label{sec-evaluation}

This section evaluates the performance, accuracy, and efficiency of Bit-Cigma in comparison to state-of-the-art bit-sparsity-aware accelerators.

\begin{table}[t]
  \centering
  \caption{Specifications of Bit-Cigma core configurations.}
  \label{tab-evalu-Configuration}
  \scriptsize
  \input{tab/tab-evalu-Configuration.tex}
\end{table}

\subsection{Experimental Setup}\label{sec-evaluation-setup}

A cycle-accurate simulator in C++ was developed to model the Bit-Cigma PE array, enabling the evaluation of performance and accuracy by simulating execution cycles and dot product results of Bit-Cigma cores.

Bit-Cigma cores were implemented in Verilog HDL with various configurations.
Using 65 nm technology (SS corner, 125℃, 1.08V), logic synthesis and netlist area evaluations were conducted using Synopsys Design Compiler v2018.06.
Power consumption was assessed using Synopsys Prime Power v2018.06 on the same technology library (TT corner, 25℃, 1.2V), based on toggle rates collected from netlist simulations using Synopsys VCS v2018.09.
Each configuration was tested on ten thousand matrices randomly selected from benchmarks corresponding to the datatype.
All memory components within Bit-Cigma preprocessors and PEs are register arrays.

\textbf{Configurations:}
Bit-Cigma cores were implemented with QINT16, FP16, BF16, and FP32 configurations as specified in Tab. \ref{tab-evalu-Configuration}.
These configurations were chosen for their relevance to various computational needs.
FP32 and FP16 are standard IEEE-754 formats for general computing, while BF16, a low-precision variant of FP32, is commonly used in AI models.
QINT16 is a quantized type applied in both AI and general applications.
To maintain a segment size of 64 data elements, the 16-bit configurations use 16 lanes with a check window size of 4 data elements, while the FP32 configuration uses 32 lanes with a check window size of 2 data elements.
All configurations employ the W bit \& A value strategy for sparsity exploitation, the ripple-carry encoder for CCSD conversion, and the round-toward-the-nearest rounding in the trailing stage.
To meet timing requirements, pipelines were inserted into the accumulator using retiming tools.

\textbf{Baselines:}
Bit-Cigma was compared against several state-of-the-art bit-sparsity-aware accelerators \cite{Pragmatic2017, Tactical2019, Tetris2018, DWP2022, Bitlet2021}, which have demonstrated superior performance and efficiency over traditional bit-parallel and value-sparsity-aware designs.
For QINT benchmarks, Tetris \cite{Tetris2018} (true form), Bit-Pragmatic \cite{Pragmatic2017} (radix-4 Booth code), Bit-Tactical \cite{Tactical2019} (radix-4 Booth code), and DWP/DWP-intra \cite{DWP2022} (CSD) served as baselines, all configured for QINT16 with 16 parallel lanes.
For FP benchmarks, Bitlet \cite{Bitlet2021} was used as a baseline, configured for FP32 with 24 parallel lanes.

\begin{table}[t]
  \centering
  \caption{Benchmark DNN models used for evaluation.}
  \label{tab-evalu-Benchmark}
  \scriptsize
  \input{tab/tab-evalu-Benchmark.tex}
\end{table}

\textbf{Benchmarks:}
Although Bit-Cigma is designed for generic matrix multiplication, most baselines focus on DNN acceleration.
Therefore, well-known DNN models were selected as benchmarks, as detailed in Tab. \ref{tab-evalu-Benchmark}.
For QINT evaluations, pre-trained AlexNet\cite{Alexnet2012}, VGG-16\cite{Vgg2015}, ResNet-152\cite{Resnet2016}, and MobileNet-v3\cite{MobileNet2019} were used, all quantized to INT16/Q12.
For FP evaluations, pre-trained Transformer\cite{Transformer2017}, converted to FP16, BF16, and FP32, were used.

\textbf{Test Data:}
For simulator-based evaluations, 800 images randomly selected from the Flickr-8k dataset and 300 records randomly selected from the WMT-14 dataset were used to generate the test data for QINT and FP benchmarks, respectively.
All supported operations in DNNs, including convolution, deconvolution, linear layer, etc., were converted to matrix multiplications.
Evaluations were conducted on all matrices generated for all selected images or records to ensure representative results.

\begin{figure}[t]
  \centering
  \includegraphics[scale=0.8]{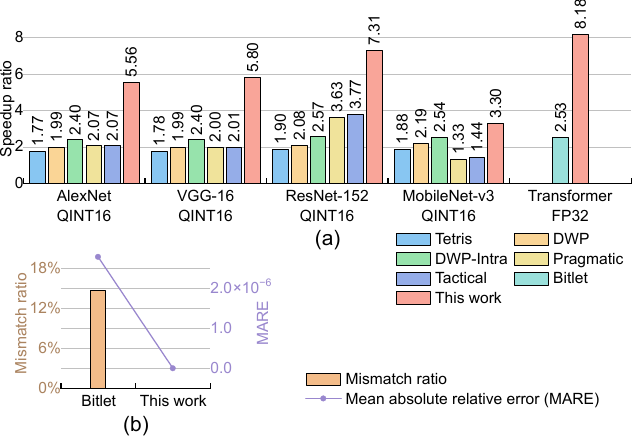}
  \caption{Comparison with bit-sparsity-aware accelerators. (a) Performance comparison based on computing cycles for a single-PE model (higher is better). (b) Accuracy comparison (lower is better). The left y-axis shows the mismatch ratio and the right y-axis presents the mean absolute relative error (MARE) across all mismatch points.}
  \label{fig-evalu-Compare}
\end{figure}

\subsection{Performance}\label{sec-evaluation-performance}

To demonstrate the advantages of acceleration using CCSD, Fig. \ref{fig-evalu-Compare}(a) presents the speedup ratio, measured by average clock cycles required to complete all matrix multiplications on single-PE models.
For QINT benchmarks, Bit-Cigma outperforms Tetris, DWP, DWP-intra, Pragmatic, and Tactical, achieving up to 3.85×, 3.50×, 2.84×, 2.91×, and 2.89× relative speedup, respectively.
For FP32 benchmarks, Bit-Cigma is 3.23× faster than Bitlet.
Across all workloads, Bit-Cigma consistently demonstrates significantly higher performance.

However, all sparsity-aware accelerators exhibit limited speedup on MobileNet-V3.
This is due to the low sparsity of MobileNet-V3, as shown in Fig. \ref{fig-motiv-CodeDensity}, making it less suitable for sparsity-focused accelerators.
Additionally, most vectors in MobileNet-V3 contain fewer than 16 elements, insufficient to fully utilize the boosting lanes, which diminishes the effectiveness of bit-serial designs including Bit-Cigma.
Despite these challenges, Bit-Cigma achieves a 3.3× speedup, outperforming other architectures by factors ranging from 1.50× to 2.49×.
This highlights the robust general applicability of the Bit-Cigma architecture with the CCSD technique.

The performance boost of Bit-Cigma primarily stems from the CCSD technique, which enhances bit-sparsity and doubles the effective check window size.
The advantages of CCSD on performance are further discussed in Sec. \ref{sec-sensitivity-encoding}.

\subsection{FP Accuracy}\label{sec-evaluation-accuracy}

To assess the impact of exponent matching on the accuracy, Fig. \ref{fig-evalu-Compare}(b) compares the mismatch ratio across all test matrices and the mean absolute relative error (MARE) across all mismatch points.
As described in Sec. \ref{sec-evaluation-setup}, computing errors were evaluated on matrices derived from the Transformer model, ensuring the results reflect practical usage scenarios.
The mismatch ratio offers a clearer indication of accuracy loss, as errors may be obscured by the redundant precision of FP32 in DNNs.
Unlike Bitlet, where mismatches are unavoidable, Bit-Cigma eliminates computing errors.

This error elimination is a result of our segmented exponent matching strategy, which aligns partial products without discarding essential bits from operands, truncating only the least significant bits in some intermediate values.
The partial product width is sufficient to ensure that any errors are covered by quantization when converting to standard FP formats.
Consequently, values with error after quantization are identical to values without error.
This ensures that Bit-Cigma produces dot products consistent with FP-represented true values.

\begin{figure}[t]
  \centering
  \includegraphics[scale=0.8]{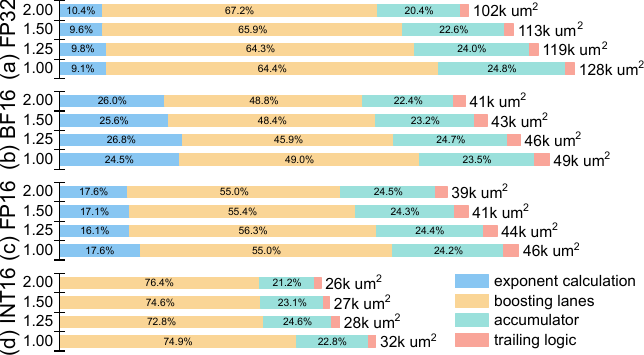}
  \caption{Area breakdown of Cigma-PE at different clock periods (ns).}
  \label{fig-evalu-AreaBreakdown}
\end{figure}

\begin{figure}[t]
  \centering
  \includegraphics[scale=0.8]{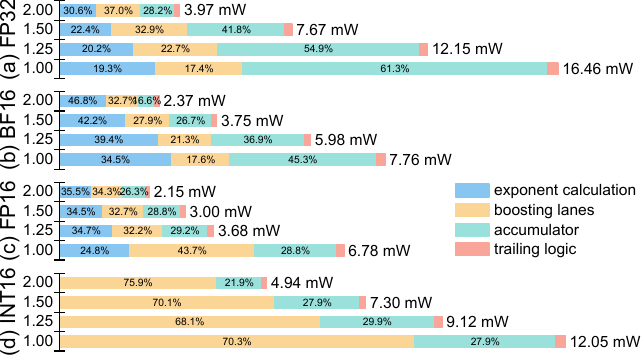}
  \caption{Power breakdown of Cigma-PE at different clock periods (ns).}
  \label{fig-evalu-PowerBreakdown}
\end{figure}

\begin{table}[t]
  \centering
  \caption{Area and power of 65 nm implementations of Cigma-STPR and Cigma-DYPR at different clock frequencies.}
  \label{tab-evalu-PrepAreaPower}
  \scriptsize
  \input{tab/tab-evalu-PrepAreaPower.tex}
\end{table}

\subsection{Area and Power}\label{sec-evaluation-areapower}

Figs. \ref{fig-evalu-AreaBreakdown} and \ref{fig-evalu-PowerBreakdown} present the area and power breakdowns of Bit-Cigma PEs at different clock frequencies.
At 1 GHz, an FP32 PE consumes 16.46 mW with an area of 0.128 mm², while an INT16 core consumes 12.05 mW and occupies 0.032 mm².
The BF16 and FP16 cores consume 7.76 mW and 6.78 mW, with areas of 0.049 mm² and 0.046 mm², respectively. 

Boosting lanes and accumulators are the primary contributors to area and power consumption.
The area of boosting lanes is two to three times that of the accumulator, varying across configurations.
Power consumption of the INT16 configuration is predominantly driven by large boosting lane check windows, whereas accumulators consume most power in FP configurations due to wide partial products.

The area and power required for exponent calculation are minimal, demonstrating the efficiency of the 2-stage exponent matching strategy in reducing circuit overhead.
In the FP32 configuration, exponent calculation accounts for 9\% of the area and 20\% of the power, a significant improvement over Bitlet, where these figures are 33\% and 66\%, respectively.

As clock frequency increases, circuit area grows slightly, while power consumption escalates exponentially.
However, the proportional distribution of area and power among the different modules remains relatively stable.
Lowering the clock frequency benefits the accumulator, as the compiler can better optimize mathematical logic.
Nevertheless, due to the logic loop in the boosting lane, our implementation is optimized for high frequencies, making frequency reduction less beneficial.

The area and power consumption of STPR and DYPR are listed in Tab. \ref{tab-evalu-PrepAreaPower}.
At 1 GHz, an FP32 STPR consumes 0.08 mW with an area of 0.03 mm², while an FP32 DYPR consumes 0.2 mW with an area of 0.016 mm².
The preprocessor costs of other configurations are approximately half that of the FP32 configuration.
Compared to the PE, the area and power consumption of both STPR and DYPR are negligible.
The constant area of the STPR across different clock frequencies indicates that the CCSD encoder circuit is sufficiently fast.

\begin{figure}[t]
  \centering
  \includegraphics[scale=0.8]{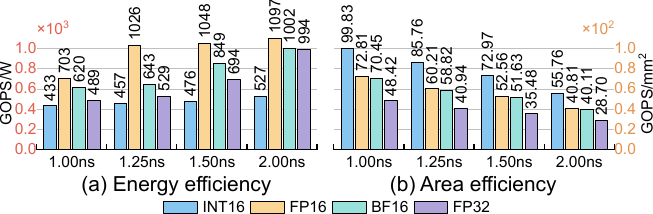}
  \caption{The energy and area efficiency of Bit-Cigma single-PE models at different clock frequencies.}
  \label{fig-evalu-Efficiency}
\end{figure}

\begin{table*}[t]
  \centering
  \caption{Comparison of area and energy efficiency with bit-sparsity-aware accelerators using 16×16 arrays, excluding SRAMs.}
  \label{tab-evalu-CircuitCompare}
  \scriptsize
  \input{tab/tab-evalu-CircuitCompare.tex}
\end{table*}

\subsection{Efficiency}\label{sec-evaluation-efficiency}

The area and energy efficiencies of a single-core Bit-Cigma implementation are depicted in Fig. \ref{fig-evalu-Efficiency}.
As the operating frequency increases, a trade-off emerges: energy efficiency decreases while area efficiency improves, reflecting the area and power changes.
Among all configurations, the FP16 mode consistently achieves the highest energy efficiency, attributable to its simpler mathematical logic components.

Tab. \ref{tab-evalu-CircuitCompare} compares the 16×16 PE array of Bit-Cigma against other designs operating at 1 GHz.
INT16 Bit-Cigma exhibits 10.7× and 3.4× higher area efficiency compared to Pragmatic and Tactical, respectively, and 13.2× greater energy efficiency over Pragmatic.
FP32 Bit-Cigma is 6.1× more efficient in area and 15.3× more efficient in energy compared to Bitlet.
The superior efficiency of Bit-Cigma stems from the synergistic effects of CCSD encoding and segmented dot product techniques.
CCSD encoding improves performance without significantly increasing circuit area.
The segmented dot product approach minimizes memory requirements within Bit-Cigma preprocessors and PEs, streamlines exponent-related logic, and boosts throughput, resulting in reduced costs and improved performance.
Additionally, the limited flexibility in supporting variable data widths within the same circuit allows for targeted optimizations in zero-skip logic and arithmetic circuits, further enhancing efficiency.

Tab. \ref{tab-evalu-CircuitCompare} also contrasts the efficiency of Bit-Cigma arrays at different frequencies.
Similar to single-PE models, higher frequencies lead to increased area efficiency but decreased energy efficiency, as the cost of PEs dominates, while the cost of STPR and DYPR remains minimal.

%% file: tab/tab-evalu-Configuration.tex
\begin{tabular}{ccccc}
\toprule
                             & INT16 & FP16  & BF16  & FP32 \\
\midrule     
lane number $L$              & 16    & 16    & 16    & 32   \\
check window size (data)     & 4     & 4     & 4     & 2    \\
check window size (bits)     & 32    & 24    & 20    & 26   \\
partial product width (bits) & 32    & 36    & 36    & 56   \\
\midrule
pipeline depth @ 1 GHz       & 1     & 1     & 2     & 3    \\
pipeline depth @ 800 MHz     & 1     & 0     & 1     & 3    \\
pipeline depth @ 667 MHz     & 0     & 0     & 1     & 2    \\
pipeline depth @ 500 MHz     & 0     & 0     & 1     & 1    \\
\midrule
segment size            & \multicolumn{4}{c}{64 data}                  \\
sparsity-aware strategy & \multicolumn{4}{c}{W bit \& A value}         \\
CCSD encoder scheme     & \multicolumn{4}{c}{ripple-carry}             \\
accumulator buffer      & \multicolumn{4}{c}{4 partial results}        \\
rounding attribute      & \multicolumn{4}{c}{round toward the nearest} \\
\bottomrule
\end{tabular}

%% file: tab/tab-evalu-Benchmark.tex
\begin{tabular}{cccc}
\toprule
  & Model & Training Dataset & Test Dataset \\
\midrule
\multirow{4}{*}{QINT}
  & AlexNet \cite{Alexnet2012} & ILSVRC-2010 & \multirow{4}{*}{Flickr-8k, 800 images} \\
  & VGG-16 \cite{Vgg2015} & ILSVRC-2012 & \\
  & ResNet-152 \cite{Resnet2016} & ImageNet & \\
  & MobileNet-v3 \cite{MobileNet2019} & ImageNet & \\
\midrule
\multirow{1}{*}{FP}
  & Transformer \cite{Transformer2017} & C4 & WMT-14, 300 records \\
\bottomrule
\end{tabular}

%% file: tab/tab-evalu-PrepAreaPower.tex
\setlength{\tabcolsep}{4pt}
{
\centering
\begin{tabular}{c@{}ccccccccc}
\toprule
      & Freq. & \multicolumn{4}{c}{Cigma-STPR} & \multicolumn{4}{c}{Cigma-DYPR} \\
\cmidrule(l{4pt}r{4pt}){3-6} \cmidrule(l{4pt}r{4pt}){7-10}
      & (MHz) & INT16 & FP16  & BF16  & FP32  & INT16 & FP16  & BF16  & FP32  \\
\midrule
\begin{tabular}[c]{@{}c@{}}
  Area \\
  (mm²)
\end{tabular}
      & \begin{tabular}[c]{@{}c@{}}
          Up to \\
          1000
        \end{tabular}
              & 0.013 & 0.012 & 0.012 & 0.025 & 0.008 & 0.008 & 0.008 & 0.016 \\
\midrule
      & 1000  & 0.041 & 0.036 & 0.034 & 0.079 & 0.102 & 0.108 & 0.110 & 0.205 \\
Power & 800   & 0.033 & 0.029 & 0.028 & 0.063 & 0.082 & 0.087 & 0.088 & 0.164 \\
(mW)
      & 667   & 0.027 & 0.024 & 0.023 & 0.053 & 0.068 & 0.072 & 0.074 & 0.137 \\
      & 500   & 0.021 & 0.018 & 0.017 & 0.040 & 0.051 & 0.054 & 0.055 & 0.103 \\
\bottomrule
\end{tabular}
}

%% file: tab/tab-evalu-CircuitCompare.tex
\begin{threeparttable}
\begin{tabular}{rccccccccccc}
\toprule
& Pragmatic\cite{Pragmatic2017} & Tactical\cite{Tactical2019} & \multicolumn{4}{c}{This work} & Bitlet\cite{Bitlet2021} & \multicolumn{4}{c}{This work} \\
& QINT16 & QINT16 & \multicolumn{4}{c}{QINT16} & FP32 & \multicolumn{4}{c}{FP32} \\
\midrule
Technology node & 65 nm & 65 nm & \multicolumn{4}{c}{65 nm} & 65 nm & \multicolumn{4}{c}{65 nm} \\
\cmidrule(lr){4-7} \cmidrule(lr){9-12}
Clock frequancy (MHz) & (1000)\tnote{*} & 1000  & 1000 & 800  & 667  & 500  & 1000  & 1000  & 800   & 667   & 500   \\
Array area (mm²)      & 68.06           & 22.97 & 8.50 & 7.57 & 7.31 & 7.08 & 46.40 & 33.54 & 31.21 & 29.64 & 26.78 \\
Array power (W)       & 32.05           & --    & 3.09 & 2.34 & 1.87 & 1.27 & 14.64 & 4.22  & 3.11  & 1.97  & 1.02  \\
\cmidrule(lr){2-7} \cmidrule(lr){8-12}
Performance (GOPS or GFLOPS)   & 541   & 558   & 1352   & 1082   & 901    & 676    & 648   & 2095   & 1676   & 1397   & 1047    \\
Area efficiency (G(FL)OPS/mm²) & 15.60 & 46.79 & 159.00 & 142.85 & 123.28 & 95.45  & 10.26 & 62.46  & 53.70  & 47.12  & 39.11   \\
Energy efficiency (G(FL)OPS/W) & 33.15 & --    & 437.88 & 462.89 & 481.68 & 534.09 & 32.52 & 496.78 & 538.24 & 710.10 & 1027.55 \\
\bottomrule 
\end{tabular}
\begin{tablenotes}
  \item[*] The clock frequency of Pragmatic is not reported. We assume a 1 GHz clock to match other works, which should represent the best-case scenario of Pragmatic.
\end{tablenotes}
\end{threeparttable}

%% file: tex/Sensitivity.tex
\section{Sensitivity Studies}\label{sec-sensitivity}

This section examines the sensitivity of key design parameters to assess their impact on performance and accuracy.

\begin{figure}[t]
  \centering
  \includegraphics[scale=0.8]{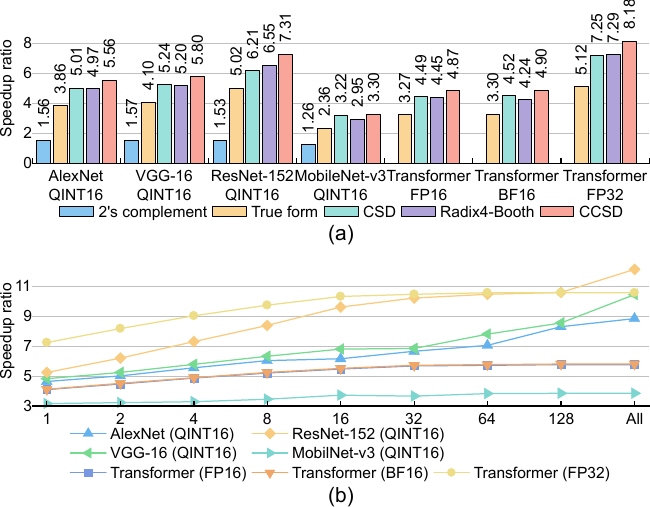}
  \caption{Sensitivity studies on speedup ratios for (a) Encoding methods and (b) Check window size. Higher is better.}
  \label{fig-evalu-ExplorePerf}
\end{figure}

\subsection{Encoding Methods}\label{sec-sensitivity-encoding}

The encoding method determines the bit-sparsity, leading to varying performance outcomes.
Fig. \ref{fig-evalu-ExplorePerf}(a) shows the speedup ratios of modified Bit-Cigma cores across various encoding methods.
CCSD consistently delivers superior performance, achieving up to 4.8×, 1.6×, 1.2×, and 1.2× improvements over 2's complement, True form, CSD, and radix-4 Booth code, respectively.
This trend aligns with the bit-sparsity patterns observed in Fig. \ref{fig-motiv-CodeDensity}.
MobileNet-V3 is an exception, as its vectors do not fully utilize the Bit-Cigma core.

The operating cycle reduction with CCSD can be attributed to three factors: efficient representation of negative values, optimized sparsity, and doubled effective check window size.
The CCSD encoder enables these advantages with negligible cost.
Although CCSD outperforms other encodings ranging from 10\% to 20\% across most benchmarks, these gains are achieved without additional circuit overhead, making CCSD an efficient and practical choice.

\subsection{Check Window Size}\label{sec-sensitivity-checkwindow}

Fig. \ref{fig-evalu-ExplorePerf}(b) illustrates the relationship between the speedup ratio and check window size, assuming an unlimited segment size.
Performance improves as the check window size grows, enhancing the capacity of the bit-sparsity boost engine to minimize synchronization delays between fast and slow lanes.
For some benchmarks, speedup plateaus as vector lengths become insufficient to fully utilize the boost engines beyond a certain check window size.

While larger check windows enhance performance, they demand higher SPAD access bandwidth.
In practice, the SPAD can cache an entire segment, with the maximum check window size constrained by the segment size.
By exposing all data in the SPAD to the PEs, Bit-Cigma achieves optimal performance at the configured segment size.
To mitigate circuit and layout complexity, all Bit-Cigma configurations limit the segment size to 64 data.

\begin{figure}[t]
  \centering
  \includegraphics[scale=0.8]{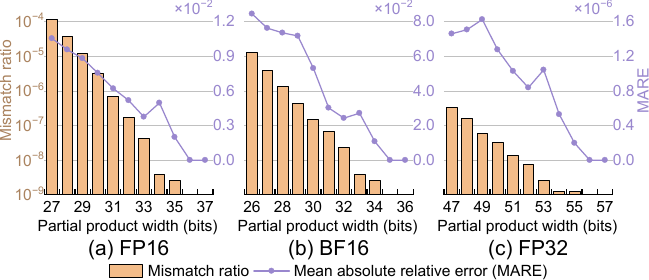}
  \caption{Accuracy-related sensitivity studies on partial product width. Lower is better. (a) FP16, (b) BF16, and (c) FP32. The left y-axis shows the mismatch ratio (logarithmic scale), and the right y-axis presents the mean absolute relative error (MARE) across all mismatch points.}
  \label{fig-evalu-ProdWidth}
\end{figure}

\subsection{Partial Product Width}\label{sec-sensitivity-productwidth}

Partial product width is critical for determining arithmetic precision, as Bit-Cigma performs exponent matching on partial products.
Fig. \ref{fig-evalu-ProdWidth} evaluates FP computing accuracy by mismatch ratio and mean absolute relative error (MARE) across mismatch points.

As partial product width increases, the mismatch ratio decreases exponentially, and relative error tends to decline.
When the partial product width is sufficiently large, mismatches are eliminated, as discussed in Sec. \ref{sec-evaluation-accuracy}.

\begin{figure*}[t]
  \centering
  \includegraphics[scale=0.8]{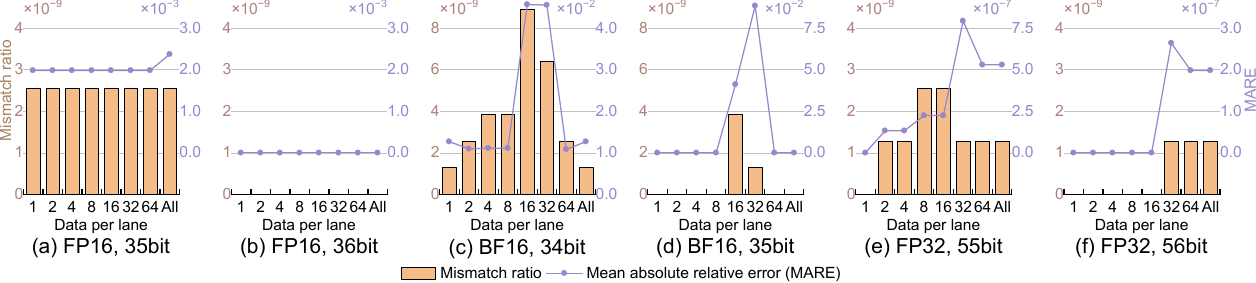}
  \caption{Accuracy-related sensitivity studies on segment size. Lower is better. Data type and partial product width are (a) FP16 with 35-bit, (b) FP16 with 36-bit, (c) BF16 with 34-bit, (d) BF16 with 35-bit, (e) FP32 with 55-bit, and (f) FP32 with 56-bit. Here, (a), (c), and (e) are on insufficient partial product width, while (b), (d), and (f) are on sufficient partial product width. The left y-axis shows the mismatch ratio and the right y-axis presents the mean absolute relative error (MARE) across all mismatch points.}
  \label{fig-evalu-SegSize}
\end{figure*}

\subsection{Segment Size}\label{sec-sensitivity-segmentsize}

The segmented approach performs FP exponent matching on each segment, and variations in segment size can affect the final result by altering the discarded essential digits.
Fig. \ref{fig-evalu-SegSize} displays the impact of segment size on dot product accuracy, comparing accuracy across different segment sizes with both sufficient and insufficient partial product widths.

In general, accuracy tends to decrease as segment size increases.
Even when partial product widths are adequate to eliminate errors with the configured segment size, larger segments can introduce errors due to the maximum points.
A maximum point holds the global maximum exponent sum within a vector pair.
Larger segment sizes are more prone to such points, since they cause a substantial shift in the second-stage exponent matching, resulting in the abrupt discarding of a large number of essential bits.
In such cases, the discarded bits cannot be canceled out, resulting in inaccuracies.
Conversely, smaller segment sizes are less affected by these maximum points, as it allows for more gradual second-stage exponent matching.
This gradual shift permits the cancellation of discarded bits.

For BF16 and FP32 configurations, accuracy appears to improve when the segment size becomes excessively large.
However, this improvement arises from the cancellation of positive and negative discarded bits, leading to quantization effects during conversion to standard FP formats, rather than from the segmented exponent matching itself.

Overall, while segment size does affect accuracy, its impact is less significant than that of partial product width.
Thus, partial product width remains the primary determinant of accuracy.
The segmented approach provides a feasible methodology for processing long vectors with limited hardware resources, offering additional benefits.

%% file: tex/ReatedWork.tex
\section{Related Work}\label{sec-relatedwork}

Accelerating matrix multiplication is a critical focus in the development of DNN hardware accelerators, leading to the emergence of various strategies.
Sparsity-aware acceleration techniques, which eliminate operations associated with zeros, have evolved from exploiting value sparsity to addressing bit-level inefficiencies.

With the bit-serial approach, Pragmatic \cite{Pragmatic2017} skips zero bits in activations, and Tactical \cite{Tactical2019} further utilizes the zero values in weights.
Laconic \cite{Laconic2019} leverages bit-sparsity in both weights and activations.
These approaches utilize radix-4 Booth encoding to reduce the number of significant digits.

Bit-interleaving techniques, such as Tetris \cite{Tetris2018} and DWP \cite{DWP2022}, focus on exploiting bit-sparsity in weights.
DWP incorporates CSD representation for better sparsity, concentrating essential bits into less significant locations.
However, the bit-interleaving approach typically performs best when bits at locations have a similar likelihood of being essential.
This characteristic limits the performance gains of DWP, which do not fully align with the sparsity reduction achieved by CSD encoding.
To address workload imbalances, DWP-intra \cite{DWP2022}, a variant of DWP, redistributes the load from the sparsest to the densest lanes, enhancing overall performance.

Bitlet \cite{Bitlet2021} introduces a floating-point (FP) dot product core that employs a bit-interleaving approach combined with an exponent matching strategy where weights handle all shifting operations.
While this method can discard many essential bits, leading to significant computational errors, these errors may be tolerable in certain DNN applications.
However, performing exponent matching on weights increases the spatiotemporal dependency between exponent and mantissa processing, causing the processing element (PE) to stall until exponent operations are completed.

Research on CSD encoding techniques dates back to earlier years, with notable contributions from Ruiz et al. \cite{Ruiz2011}, Guo et al. \cite{Guo2011}, and Tanaka et al. \cite{Tanaka2016}, who have developed encoding circuits for various CSD binary codes.
These works employ $N$ CSD digits to represent an $N$-bit integer, using two binary bits per CSD digit.
Although these studies optimize the carry chain through different methods, they all rely on an $N$-bit carry chain to encode $N$-bit data.

%% file: tex/Conclusion.tex
\section{Conclusion}\label{sec-conclusion}

This paper presents Bit-Cigma, a bit-sparsity-aware hardware accelerator designed for generic matrix multiplication, focusing on high accuracy, performance, and efficiency.
The core innovation lies in the Compact CSD (CCSD) encoding technique, which optimizes bit-level sparsity through a fully on-chip process with minimal cost.
Additionally, the segmented dot product approach enables efficient processing of large matrices and vectors, which enhances exponent matching and data reuse by treating matrices as blocks and segments.
The CCSD technique allows Bit-Cigma to use standard FP or QINT formats for memory access, achieving significant performance gains without incurring additional memory access overhead.
Furthermore, by supporting standard exchange formats, Bit-Cigma can accelerate original DNNs without requiring software-based preliminaries.
When evaluated on DNN models using FP32 data, Bit-Cigma demonstrates a 3.2× performance improvement, 6.1× area efficiency, and 15.3× energy efficiency compared to the state-of-the-art Bitlet.
We believe the CCSD technique and segmented dot product approach open new avenues for research and innovation in general-purpose acceleration.

%% file: refs.bib
@inproceedings{Pragmatic2017,
  title     = {Bit-pragmatic deep neural network computing},
  author    = {Albericio, Jorge and Delm{\'a}s, Alberto and Judd, Patrick and Sharify, Sayeh and O'Leary, Gerard and Genov, Roman and Moshovos, Andreas},
  booktitle = {Proceedings of the 50th Annual IEEE/ACM International Symposium on Microarchitecture},
  pages     = {382--394},
  year      = {2017},
  month     = {Oct},
  doi       = {10.1145/3123939.3123982}
}

@inproceedings{Tetris2018,
  title     = {Tetris: Re-architecting Convolutional Neural Network Computation for Machine Learning Accelerators},
  author    = {Lu, Hang and Wei, Xin and Lin, Ning and Yan, Guihai and Li, Xiaowei},
  booktitle = {2018 IEEE/ACM International Conference on Computer-Aided Design (ICCAD)},
  pages     = {1--8},
  year      = {2018},
  month     = {Nov},
  doi       = {10.1145/3240765.3240855}
}

@inproceedings{Tactical2019,
  title     = {Bit-tactical: A software/hardware approach to exploiting value and bit sparsity in neural networks},
  author    = {Delmas Lascorz, Alberto and Judd, Patrick and Stuart, Dylan Malone and Poulos, Zissis and Mahmoud, Mostafa and Sharify, Sayeh and Nikolic, Milos and Siu, Kevin and Moshovos, Andreas},
  booktitle = {Proceedings of the Twenty-Fourth International Conference on Architectural Support for Programming Languages and Operating Systems},
  pages     = {749--763},
  year      = {2019},
  month     = {Apr},
  doi       = {10.1145/3297858.3304041}
}

@inproceedings{Laconic2019,
  title     = {Laconic deep learning inference acceleration},
  author    = {Sharify, Sayeh and Lascorz, Alberto Delmas and Mahmoud, Mostafa and Nikolic, Milos and Siu, Kevin and Stuart, Dylan Malone and Poulos, Zissis and Moshovos, Andreas},
  booktitle = {2019 ACM/IEEE 46th Annual International Symposium on Computer Architecture (ISCA)},
  pages     = {304--317},
  year      = {2019},
  month     = {June},
  doi       = {10.1145/3307650.3322255}
}

@article{DWP2022,
  title   = {Deeper Weight Pruning Without Accuracy Loss in Deep Neural Networks: Signed-Digit Representation-Based Approach},
  author  = {Ahn, Byungmin and Kim, Taewhan},
  journal = {IEEE Transactions on Computer-Aided Design of Integrated Circuits and Systems},
  volume  = {41},
  number  = {3},
  pages   = {656--668},
  year    = {2022},
  month   = {Mar},
  doi     = {10.1109/TCAD.2021.3064914}
}

@article{Balance2024,
  title   = {Bit-Balance: Model-Hardware Codesign for Accelerating NNs by Exploiting Bit-Level Sparsity}, 
  author  = {Sun, Wenhao and Zou, Zhiwei and Liu, Deng and Sun, Wendi and Chen, Song and Kang, Yi},
  journal = {IEEE Transactions on Computers}, 
  volume  = {73},
  number  = {1},
  pages   = {152--163},
  year    = {2024},
  month   = {Jan},
  doi     = {10.1109/TC.2023.3324477}
}

@article{Jing2023,
  title   = {Exploiting bit sparsity in both activation and weight in neural networks accelerators},
  author  = {Jing, Naifeng and Zhang, Zihan and Sun, Yongshuai and Liu, Pengyu and Chen, Liyan and Wang, Qin and Jiang, Jianfei},
  journal = {Integr. VLSI J.},
  volume  = {88},
  number  = {C},
  pages   = {400--409},
  year    = {2023},
  month   = {Jan},
  doi     = {10.1016/j.vlsi.2022.09.008},
}

@InProceedings{BitHist2023,
  title     = {BitHist: A Precision-Scalable Sparse-Awareness DNN Accelerator Based on Bit Slices Products Histogram},
  author    = {Meng, Zhaoteng and Xiao, Long and Gao, Xiaoyao and Li, Zhan and Shu, Lin and Hao, Jie},
  booktitle = {Euro-Par 2023: Parallel Processing},
  pages     = {289--303},
  year      = {2023},
  month     = {Aug},
  doi       = {10.1007/978-3-031-39698-4_20}
}

@inproceedings{Bitlet2021,
  title     = {Distilling Bit-Level Sparsity Parallelism for General Purpose Deep Learning Acceleration},
  author    = {Lu, Hang and Chang, Liang and Li, Chenglong and Zhu, Zixuan and Lu, Shengjian and Liu, Yanhuan and Zhang, Mingzhe},
  booktitle = {MICRO-54: 54th Annual IEEE/ACM International Symposium on Microarchitecture},
  pages     = {963--976},
  year      = {2021},
  month     = {Oct},
  doi       = {10.1145/3466752.3480123}
}

@inproceedings{BEM2022,
  title     = {BEM: Bit-level Sparsity-aware Deep Learning Accelerator with Efficient Booth Encoding and Weight Multiplexing}, 
  author    = {Gao, Yunhung and Zhang, Kevin and Jia, Song},
  booktitle = {2022 IEEE 4th International Conference on Circuits and Systems (ICCS)}, 
  pages     = {187--194},
  year      = {2022},
  month     = {Sep},
  doi       = {10.1109/ICCS56666.2022.9936588},
}

@inproceedings{Mortar2023,
  title     = {Mortar: Morphing the Bit Level Sparsity for General Purpose Deep Learning Acceleration},
  author    = {Gao, Yunhung and Li, Hongyan and Zhang, Kevin and Yu, Xueru and Lu, Hang},
  booktitle = {Proceedings of the 28th Asia and South Pacific Design Automation Conference},
  pages     = {739--744},
  year      = {2023},
  month     = {Jan},
  doi       = {10.1145/3566097.3567868},
}

@article{NeuroFlix2022,
  title={A Deep Neural Network Training Architecture With Inference-Aware Heterogeneous Data-Type}, 
  author={Choi, Seungkyu and Shin, Jaekang and Kim, Lee-Sup},
  journal={IEEE Transactions on Computers}, 
  volume={71},
  number={5},
  pages={1216--1229},
  year={2022},
  month={May},
  doi={10.1109/TC.2021.3078316}
}

@article{FlexBlock2023,
  title={FlexBlock: A Flexible DNN Training Accelerator With Multi-Mode Block Floating Point Support}, 
  author={Noh, Seock-Hwan and Koo, Jahyun and Lee, Seunghyun and Park, Jongse and Kung, Jaeha},
  journal={IEEE Transactions on Computers}, 
  volume={72},
  number={9},
  pages={2522--2535},
  year={2023},
  month={Sep.},
  doi={10.1109/TC.2023.3253050}
}

@ARTICLE{BitletX2024,
  title={General Purpose Deep Learning Accelerator Based on Bit Interleaving}, 
  author={Chang, Liang and Lu, Hang and Li, Chenglong and Zhao, Xin and Hu, Zhicheng and Zhou, Jun and Li, Xiaowei},
  journal={IEEE Transactions on Computer-Aided Design of Integrated Circuits and Systems}, 
  volume={43},
  number={5},
  pages={1470--1483},
  year={2024},
  month={May},
  doi={10.1109/TCAD.2023.3342728}
}

@inproceedings{Hewlitt2000,
  title     = {Canonical signed digit representation for FIR digital filters},
  author    = {Hewlitt, Reid M and Swartzlantler, ES},
  booktitle = {2000 IEEE Workshop on SiGNAL PROCESSING SYSTEMS. SiPS 2000. Design and Implementation (Cat. No. 00TH8528)},
  pages     = {416--426},
  year      = {2000},
  month     = {Oct},
  doi       = {10.1109/SIPS.2000.886740}
}

@inproceedings{HWP2004,
  title     = {HWP: A new insight into canonical signed digit},
  author    = {Xu, Fei and Chang, Chip-Hong and Jong, Ching-Chuen},
  booktitle = {2004 IEEE International Symposium on Circuits and Systems (ISCAS)},
  pages     = {V--V},
  year      = {2004},
  month     = {May},
  doi       = {10.1109/ISCAS.2004.1329497}
}

@article{Ruiz2011,
  title   = {Efficient canonic signed digit recoding},
  author  = {Ruiz, Gustavo A and Granda, Mercedes},
  journal = {Microelectronics journal},
  volume  = {42},
  number  = {9},
  pages   = {1090--1097},
  year    = {2011},
  month   = {Sept},
  doi     = {10.1016/j.mejo.2011.06.006}
}

@inproceedings{Guo2011,
  title     = {A novel fast canonical-signed-digit conversion technique for multiplication},
  author    = {Guo, Rui and DeBrunner, Linda S},
  booktitle = {2011 IEEE International Conference on Acoustics, Speech and Signal Processing (ICASSP)},
  pages     = {1637--1640},
  year      = {2011},
  month     = {May},
  doi       = {10.1109/ICASSP.2011.5946812}
}

@article{Tanaka2016,
  title   = {Efficient signed-digit-to-canonical-signed-digit recoding circuits},
  author  = {Tanaka, Yuuki},
  journal = {Microelectronics Journal},
  volume  = {57},
  number  = {3},
  pages   = {21--25},
  year    = {2016},
  month   = {Nov},
  doi     = {10.1016/j.mejo.2016.09.001}
}

@inproceedings{Alexnet2012,
  title     = {ImageNet Classification with Deep Convolutional Neural Networks},
  author    = {Krizhevsky, Alex and Sutskever, Ilya and Hinton, Geoffrey E.},
  booktitle = {Proceedings of the 25th International Conference on Neural Information Processing Systems (NIPS)},
  pages     = {1097--1105},
  year      = {2012},
  month     = {Dec},
  doi       = {}
}

@inproceedings{Vgg2015,
  title     = {Very deep convolutional neural network based image classification using small training sample size},
  author    = {Liu, Shuying and Deng, Weihong},
  booktitle = {2015 3rd IAPR Asian Conference on Pattern Recognition (ACPR)},
  pages     = {730--734},
  year      = {2015},
  month     = {Nov},
  doi       = {10.1109/ACPR.2015.7486599}
}

@inproceedings{Resnet2016,
  title     = {Deep residual learning for image recognition},
  author    = {He, Kaiming and Zhang, Xiangyu and Ren, Shaoqing and Sun, Jian},
  booktitle = {Proceedings of the IEEE Conference on Computer Vision and Pattern Recognition (CVPR)},
  pages     = {770--778},
  year      = {2016},
  month     = {June},
  doi       = {10.1109/CVPR.2016.90}
}

@inproceedings{Transformer2017,
  title     = {Attention is All You Need},
  author    = {Vaswani, Ashish and Shazeer, Noam and Parmar, Niki and Uszkoreit, Jakob and Jones, Llion and Gomez, Aidan N. and Kaiser, \L{}ukasz and Polosukhin, Illia},
  booktitle = {Proceedings of the 31st International Conference on Neural Information Processing Systems},
  pages     = {6000--6010},
  year      = {2017},
  month     = {Dec},
  doi       = {}
}

@inproceedings{MobileNet2019,
  title = {Searching for MobileNetV3},
  author = {A. Howard and M. Sandler and B. Chen and W. Wang and L. Chen and M. Tan and G. Chu and V. Vasudevan and Y. Zhu and R. Pang and H. Adam and Q. Le},
  booktitle = {2019 IEEE/CVF International Conference on Computer Vision (ICCV)},
  pages = {1314--1324},
  year = {2019},
  month = {Nov},
  doi = {10.1109/ICCV.2019.00140}
}

@article{Palm2023,
  title   = {Palm: Scaling language modeling with pathways},
  author  = {Chowdhery, Aakanksha and Narang, Sharan and Devlin, Jacob and Bosma, Maarten and Mishra, Gaurav and Roberts, Adam and Barham, Paul and Chung, Hyung Won and Sutton, Charles and Gehrmann, Sebastian and others},
  journal = {Journal of Machine Learning Research},
  volume  = {24},
  number  = {240},
  pages   = {1--113},
  year    = {2023},
  month   = {},
  doi     = {}
}

@article{Sauer2023,
  title   = {Adversarial diffusion distillation},
  author  = {Sauer, Axel and Lorenz, Dominik and Blattmann, Andreas and Rombach, Robin},
  journal = {arXiv preprint arXiv:2311.17042},
  year    = {2023}
}

@article{llama2023,
  title   = {Llama 2: Open foundation and fine-tuned chat models},
  author  = {Touvron, Hugo and Martin, Louis and Stone, Kevin and Albert, Peter and Almahairi, Amjad and Babaei, Yasmine and Bashlykov, Nikolay and Batra, Soumya and Bhargava, Prajjwal and Bhosale, Shruti and others},
  journal = {arXiv preprint arXiv:2307.09288},
  year    = {2023}
}

@article{SAM2023,
  title   = {Segment anything},
  author  = {Kirillov, Alexander and Mintun, Eric and Ravi, Nikhila and Mao, Hanzi and Rolland, Chloe and Gustafson, Laura and Xiao, Tete and Whitehead, Spencer and Berg, Alexander C and Lo, Wan-Yen and others},
  journal = {arXiv preprint arXiv:2304.02643},
  year    = {2023}
}
